\documentclass[twocolumn]{aa}
\usepackage{graphicx}
\usepackage{txfonts}
\usepackage{latexsym}
\usepackage{amssymb}
\usepackage{amsfonts}
\usepackage{natbib}
\begin{document}
   \title{The mass-energy budget of the ionised outflow in NGC~7469}

   \author{A. J. Blustin
          \inst{1}\thanks{E-mail: ajb@mssl.ucl.ac.uk}
          \and
          G. A. Kriss
          \inst{2}
          \and
          T. Holczer
          \inst{3}
          \and
          E. Behar
          \inst{3}
          \and
          J. S. Kaastra
          \inst{4}
          \and
          M. J. Page
          \inst{1}
          \and
          S. Kaspi
          \inst{3,5}
          \and
          G. Branduardi-Raymont
          \inst{1}
          \and
          K. C. Steenbrugge
          \inst{6}
          }

   \offprints{A. J. Blustin}

   \institute{UCL Mullard Space Science Laboratory, Holmbury St Mary, Dorking, Surrey RH5 6NT, UK
             \and
	     Space Telescope Science Institute, 3800 San Martin Drive, Baltimore, MD 21218, USA
             \and
             Physics Department, Technion, Haifa 32000, Israel
             \and
             SRON Netherlands Institute for Space Research, Sorbonnelaan 2, 3584 CA Utrecht, Netherlands
             \and
	     School of Physics and Astronomy, Raymond and Beverley Sackler Faculty of Exact Sciences, 
	     Tel Aviv University, Tel Aviv 69978, Israel
             \and
             St. John's College Research Centre, St. John's College, University of Oxford, Oxford OX1 3JP, UK
             }

   \date{Received 5 December 2006 / Accepted 6 February 2007}

   \abstract{Although AGN feedback through ionised winds is of great importance in models of AGN/galaxy coevolution, the mass and energy output via these winds, even in the nearby universe, is poorly understood. The issue is complicated by the wide range of ionisation in the winds, which means that multiwavelength observational campaigns are required to obtain the complete picture. In this paper, we use a $\sim$ 160~ks XMM-Newton RGS spectrum to get the most accurate view yet of the ionised outflow (warm absorber) in \object{NGC~7469} as seen in X-rays, finding that there is a wide range of ionisation, with log $\xi$ in the range $\sim$ 0.5$-$3.5~erg cm s$^{-1}$, and two main velocity regimes, at 580$-$720 and 2300~km~s$^{-1}$, with the highest velocity gas being the least ionised. The total absorbing column density in the X-rays is of order 3~$\times$~10$^{21}$~cm$^{-2}$. We find that the lowest ionisation phase of the absorber is probably identical with one of the phases of the UV absorber discovered in previous studies. We show that both X-ray and UV absorbers are consistent with an origin near the base of a torus wind, where matter is being launched and accelerated. Calculating the mass outflow rate and kinetic luminosity of all the absorber phases, we demonstrate that the X-ray absorbing gas carries respectively $\sim$ 90\% and 95\% of the mass and kinetic energy output of the ionised outflow. 
   
   }

   \maketitle
%

\section{Introduction}

Ionised winds from AGN are an important and somewhat enigmatic factor in the feedback process controlling black hole and galaxy evolution. Although they have an important role in a number of popular evolutionary schemes \citep[e.g.][]{silk1998,fabian1999,granato2004,page2004}, their physical properties - such as mass outflow rate and kinetic luminosity - are not observationally well established. The situation is complicated by the fact that the winds contain gas at a wide range of ionisation levels, and so astronomers working in a single wavelength range, on either X-ray `warm absorbers' or UV `intrinsic absorbers', will receive an incomplete picture of the phenomenon.

Combined X-ray and UV spectroscopic studies of bright nearby AGN have become increasingly popular \citep[e.g.][]{mathur1994,mathur1995,collinge2001,crenshaw2003,yaqoob2003,kraemer2003}, especially in the last five years since high-resolution soft X-ray spectra from Chandra and XMM-Newton have become available. Although most of these analyses have been concerned with the low-velocity outflows in nearby Seyfert galaxies, due to the need to get good signal-to-noise in the X-rays, multiwavelength studies of high-velocity quasar outflows, like that in \object{PDS~456} \citep{obrien2005} have begun to appear. Most attention to date has focused on whether the UV and X-ray absorption lines come from the same gas, with the current conclusion being that in some cases they do, and in some cases they don't \citep[e.g.][]{blustin2005}. It has been clear for a while, however, that probably all AGN winds produce spectral signatures in both bands.

\object{NGC~7469} is a nearby (z=0.0164, \citealt{devaucouleurs1991}) Seyfert galaxy that has been the target of a number of observing campaigns by X-ray and UV observatory satellites. In recent years attention has focused on the outflow of ionised gas from its nucleus which leaves spectral traces across the X-ray and UV bands. The relationship of the different velocity and ionisation phases of the gas seen in the two wavebands, and the place of the outflow within the nuclear region, have been of particular interest. Our previous combined study of the X-ray and UV properties of the warm absorber in this source (using a 40~ks XMM-Newton observation in 2000 and a FUSE spectrum obtained a year earlier \citealt{blustin2003,kriss2003}) indicated that the X-ray absorber had a wide range of ionisation, and that the principal phase of X-ray absorption was identical with one of the two velocity/ionisation phases in the UV absorber. A study of simultaneous Chandra, FUSE and HST STIS spectra of \object{NGC~7469} \citep{scott2005}, obtained in 2002 (with some extra STIS observations from 2004), was broadly in agreement with our earlier conclusions, but neither the Chandra HETG spectrum of \citet{scott2005} nor the XMM-Newton RGS spectrum of \citet{blustin2003} had high enough signal-to-noise to satisfactorily constrain the properties of the X-ray absorber. 

At the end of 2004, the deepest ever X-ray spectroscopic observational campaign on this source was performed using the XMM-Newton RGS. In this paper we use the resulting high signal-to-noise soft X-ray spectrum, together with results from the \citet{scott2005} UV spectroscopic campaign, to estimate the relative contributions of the X-ray and UV absorbers to the overall mass outflow rate and kinetic luminosity of the AGN wind in \object{NGC~7469}. We also investigate what the new data tell us about the origin of the ionised outflow in this object.


\section{Data analysis}

\object{NGC~7469} was observed by XMM-Newton for a total of 164~ks during two long observations in November/December 2004. During these observations, EPIC-pn was operated in Small Window mode, the EPIC-MOS cameras were in Timing Uncompressed mode and the RGS was in Spectroscopy mode. The OM took long sequences of 1000~s exposures in Image mode, starting in both cases with a cycle through the V, U, UVW1, UVM2 and UVW2 filters and continuing in UVW2 until the end of the respective observations. The full observation details are listed in Table~\ref{obs_details}.

\begin{table*}
\caption{Observation Identifiers, start and end dates and times, instrument modes and total exposure times.}
\label{obs_details} 
\centering  
\begin{tabular}{lllllr}
\hline\hline 
\noalign{\smallskip}
Obs-ID & Observation start (UT) & Observation end (UT) & Instrument & Mode & Exposure (s) \\ 
\noalign{\smallskip}
\hline 
\noalign{\smallskip}
0207090101 & 2004-11-30T20:48:52 & 2004-12-01T20:43:52 & EPIC-MOS 1 & Timing Uncompressed (medium filter) & 84~502 \\ 
           &                     &            & EPIC-MOS 2 & Timing Uncompressed (medium filter) & 84~507 \\ 
           &                     &            & EPIC-PN    & Small Window (medium filter)        & 84~564 \\ 
           &                     &            & RGS 1      & Spectroscopy                        & 84~985 \\ 
           &                     &            & RGS 2      & Spectroscopy                        & 84~980 \\ 
           &                     &            & OM         & Image (V filter)                    & 1~000  \\ 
           &                     &            & OM         & Image (U filter)                    & 1~000  \\ 
           &                     &            & OM         & Image (UVW1 filter)                 & 1~000  \\ 
           &                     &            & OM         & Image (UVM2 filter)                 & 1~229  \\ 
           &                     &            & OM         & Image (UVW2 filter)                 & 57~860 \\ 
\noalign{\smallskip}
\hline 
\noalign{\smallskip}
0207090201 & 2004-12-03T01:04:39 & 2004-12-03T23:21:19  & EPIC-MOS 1 & Timing Uncompressed (medium filter) & 78~602 \\  
           &                     &            & EPIC-MOS 2 & Timing Uncompressed (medium filter) & 78~607 \\ 
           &                     &            & EPIC-PN    & Small Window (medium filter)        & 78~664 \\ 
           &                     &            & RGS 1      & Spectroscopy                        & 79~085 \\ 
           &                     &            & RGS 2      & Spectroscopy                        & 79~080 \\ 
           &                     &            & OM         & Image (V filter)                    & 1~000 \\ 
           &                     &            & OM         & Image (U filter)                    & 1~000 \\ 
           &                     &            & OM         & Image (UVW1 filter)                 & 1~000 \\ 
           &                     &            & OM         & Image (UVM2 filter)                 & 1~220 \\ 
           &                     &            & OM         & Image (UVW2 filter)                 & 53~000 \\ 
\noalign{\smallskip}
\hline 
\end{tabular}
\end{table*}

This paper deals only with the RGS, EPIC-pn and OM data. The RGS and EPIC-pn data were reduced under SAS V6.1; the RGS data were processed using \emph{rgsproc}, which extracts spectra and background-subtracts them using regions adjacent in cross-dispersion space to that containing the source. Since the response matrices generated by \emph{rgsproc} take into account the differences in effective area between the two RGS units, it is possible to combine both the RGS1 and RGS2 spectra and response matrices \citep[see][]{page2003}; both first and second order spectra were included. Remaining instrumental effects, not yet included in the calibration, were removed with the aid of residuals to power-law fits to the pure continuum source Markarian 421. The EPIC-pn data were processed first using \emph{epproc}, and then source and background spectra and lightcurves were extracted from circular extraction regions of 40'' radius. For both the RGS and EPIC-pn data, the event lists were screened to remove intervals of high proton background before spectral extraction took place. The total good time intervals for the RGS and pn were, respectively, 157~ks (a 4.4\% data loss) and 163~ks (a 0.2\% data loss).

The OM image mode photometry was done using the \emph{omichain} pipeline running under SAS V7. The resulting source lists contain count rates for each source detected, with corrections applied for various systematic effects including detector deadtime and coincidence loss. We also corrected for Galactic reddening using an E(B-V) of 0.069 obtained from the NASA Extragalactic Database (NED), and calculated monochromatic fluxes using the AB magnitude system \citep{oke1974} flux conversion factors provided in the merged source list.

Errors quoted for fitted or calculated quantities are for $\Delta \chi^{2}$ = 2 (the RMS of the $\Delta \chi^{2}$ distribution) throughout this paper.

\section{EPIC-pn spectroscopy and UV/X-ray lightcurves}

\begin{figure}
   \centering
   \includegraphics[width=6cm,angle=-90]{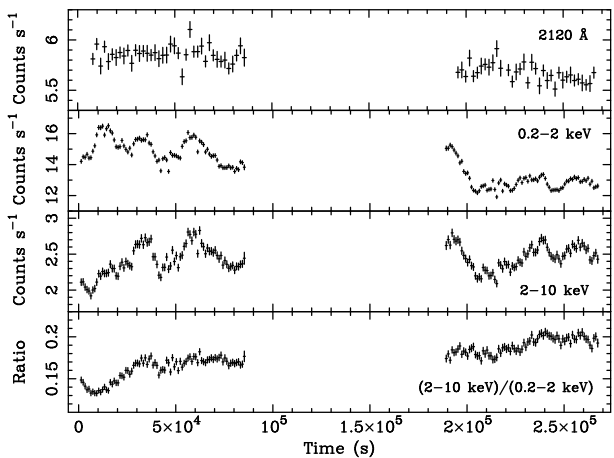}
      \caption{2120~\AA\, (OM UVW2 filter), 0.2-2~keV and 2-10~keV band pn lightcurves, with hardness ratio (2-10~keV/0.2-2~keV).
              }
         \label{pn_ltcv}
   \end{figure}

\begin{figure}
   \centering
   \includegraphics[width=6cm,angle=-90]{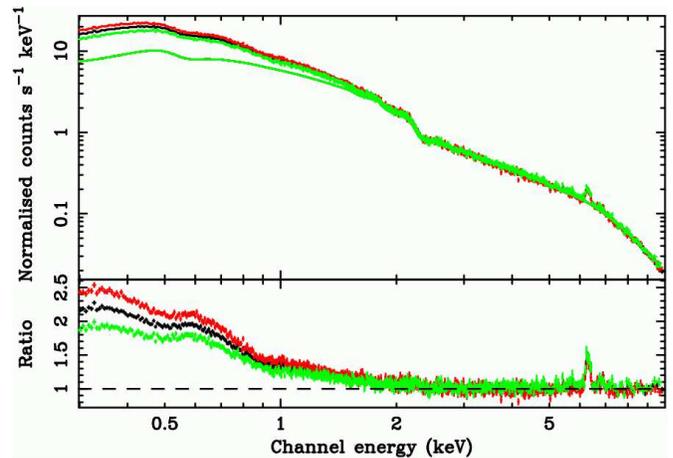}
      \caption{EPIC-pn spectra from the whole observation (black dataset; 750 counts per bin), the first (top dataset; 375 counts per bin) and second (bottom dataset; 370 counts per bin) halves of the observation with best fit power-law to the ranges 3$-$5.5~keV and 7-10~keV superimposed (continuous line). The bottom panel shows the ratio of these datasets to the model.
              }
         \label{pn_total_ratio}
   \end{figure}

The EPIC-pn spectra were used to help define the soft X-ray continuum to aid in modelling the RGS spectrum. The general shape of the pn spectrum is similar to that observed in previous XMM-Newton and Chandra observations of this object \citep{blustin2003,scott2005}; there is a high-energy power-law with a soft excess. Firstly, the lightcurves were examined for signs of spectral variability that could affect the apparent form of the spectrum. Fig.~\ref{pn_ltcv} shows the 2120~\AA\, (OM UVW2), 0.2$-$2 and 2$-$10~keV pn lightcurves alongside the 2$-$10~keV/0.2$-$2~keV hardness ratio. The spectrum hardens over the course of the observation; a comparison of pn spectra from the first and second halves of the observation (Fig.~\ref{pn_total_ratio}) reveals that although the total flux in the soft excess decreases between the two observations, the basic shape of the soft excess remained the same. Detailed analysis of the variability is presented elsewhere \citep{blustin2006}, but for our purposes here, the soft X-ray continuum shape does not seem to be seriously affected by spectral variability. 

\begin{figure}
   \centering
   \includegraphics[width=7cm]{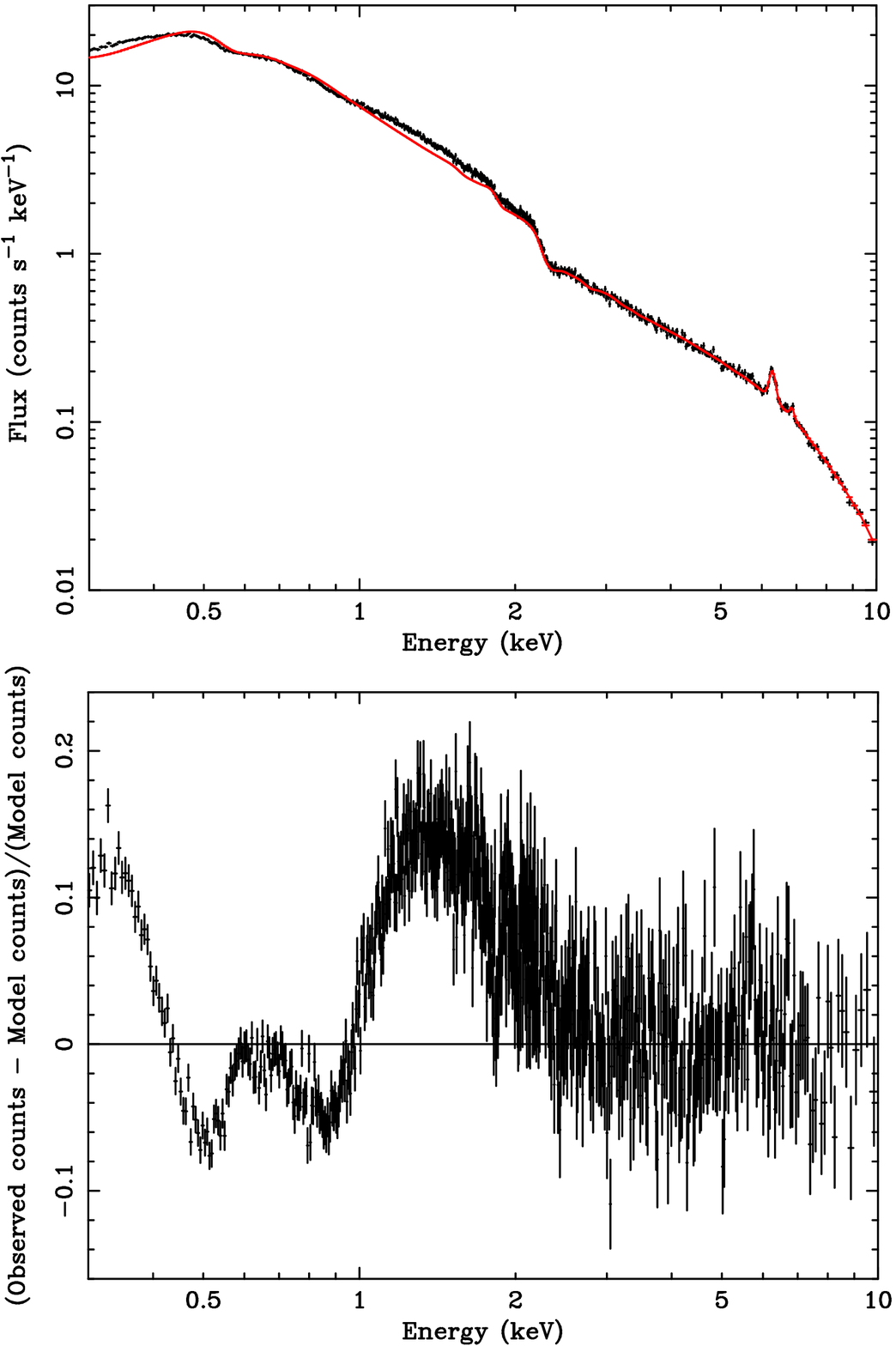}
      \caption{Top: EPIC-pn spectrum (black) with fitted model (red) comprising a Galactic-absorbed power-law, Fe~K$\alpha$ and \ion{Fe}{xxvi} lines, and a blackbody soft excess. Bottom: the residuals of the fit above; the model cannot explain the spectral complexity in the soft excess.
              }
         \label{pn_fit_chi}
   \end{figure}

In all X-ray spectral fits reported in this paper, we have incorporated neutral absorption due to our Galaxy at a column of N$_{H}$ = $4.82\times10^{20}$~cm$^{-2}$ \citep{elvis1989}. Fitting a power-law to the 3$-$10~keV range, excluding 5.5$-$7~keV, we find a slope $\Gamma$ of 1.81$\pm$0.01. In the 2000 observations, the spectral index was 1.75 \citep{blustin2002}, and in 2002 it had increased to 1.79 \citep{scott2005}. There is a Fe~K$\alpha$ line, well-fitted with a Gaussian profile, with a FWHM of 6900$\pm$1400~km~s$^{-1}$, consistent with that observed with the Chandra HETGS two years earlier \citep{scott2005} and with the first XMM-Newton observation four years earlier \citep{blustin2003}. A second weaker emission line, at a rest-frame energy of 6.97$\pm$0.03~keV, is most consistent with \ion{Fe}{xxvi}. After fixing the parameters of the hard power-law and the Fe~K$\alpha$ and \ion{Fe}{xxvi} lines, we added a blackbody component to represent the soft excess. The 0.3$-$10~keV flux was 5.7 $\times$ 10$^{-11}$~erg~s$^{-1}$; in the previous XMM-Newton observation in 2000 the flux was 4.4 $\times$ 10$^{-11}$~erg~s$^{-1}$ in this range. The resulting 0.3$-$10~keV fit is very poor, with a reduced $\chi^{2}$ of 14.35 (622 degrees of freedom). Fig.~\ref{pn_fit_chi} shows the model superimposed on the pn data, along with the residuals to the fit. The main cause of the bad fit is the inability of the model to replicate the form or spectral complexity of the soft excess, which clearly includes spectral features unresolved by the EPIC-pn. The next section describes detailed spectroscopy of these features with the RGS. 

%

%

%

\section{RGS spectroscopy}

\subsection{Line identification}

\begin{figure*}
   \centering
   \includegraphics[width=12cm]{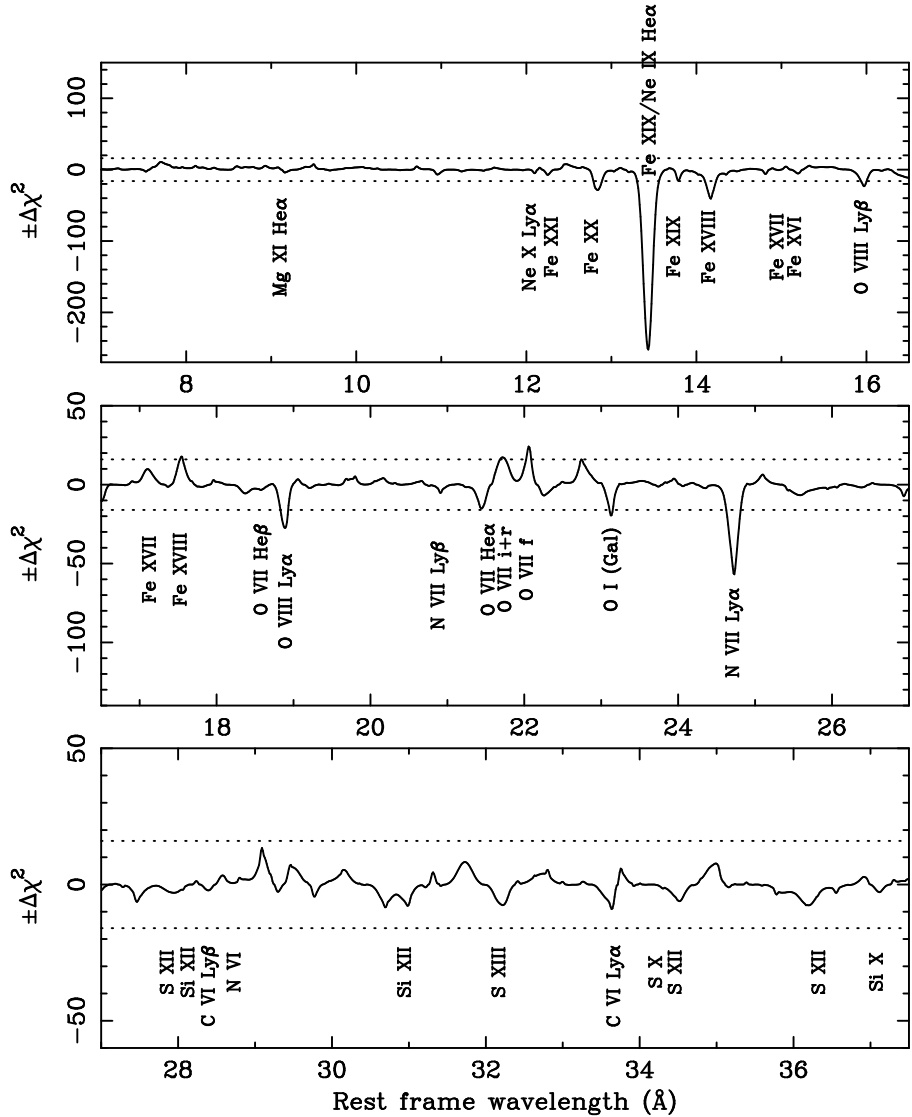}
      \caption{Plot of $\Delta\chi^{2}$ associated with fitting a Gaussian profile to features deviating from a smoothed continuum across the RGS spectrum of \object{NGC~7469}, in the rest frame of the AGN. Here a positive deviation is an emission feature, and a negative deviation implies a feature seen in absorption. $\Delta\chi^{2}$ is equivalent in this case to the square of the number of sigma significance (here a $\Delta\chi^{2}$ of 16 is equivalent to 4$\sigma$ significance for one interesting parameter; this significance level is marked in each plot with dotted lines). Possible identifications of features are labelled.
              }
         \label{chi}
   \end{figure*}

The RGS spectrum contains a number of narrow absorption and emission line features. To assess the significance of these, we used a routine which fits a Gaussian profile to spectral features deviating from a smoothed continuum. We can use the resulting plot of $\Delta\chi^{2}$ of the Gaussian fit against wavelength (Fig.~\ref{chi}) to estimate the statistical significance of the features; Table~\ref{eqws} lists the strongest absorption and emission features, significant at $>$ 99.99\% confidence (here a $|\Delta\chi^{2}|$ of 16 is equivalent to 4$\sigma$ significance for one interesting parameter).

%

\begin{table*}
\caption{Parameters of the strongest narrow absorption and emission lines in the RGS spectrum of \object{NGC~7469} (significant at $>$ 99.99\% confidence in Fig.~\ref{chi}). Columns 1 to 6: $\lambda_{rest,~meas}$, the measured wavelength in the (cosmological) rest frame in \AA\, (\ion{O}{i}$_{Gal}$ is measured in the rest frame of our Galaxy); the line identification; $\lambda_{rest,~lab}$, the theoretical rest frame wavelength of the line; $EW$, the equivalent width of the line in m\AA\, (negative values correspond to emission lines); $FWHM$, the line FWHM in km~s$^{-1}$; v$_{shift}$, the redshift or blueshift of the line in km~s$^{-1}$.}
\label{eqws}
\centering  
\begin{tabular}{c c c c c c}
\hline\hline 
$\lambda_{rest,~meas}$ & Line & $\lambda_{rest,~lab}$ & $EW$ & $FWHM$ & v$_{shift}$ \\ 
\hline   
24.73$^{+0.02}_{-0.01}$ & \ion{N}{vii} Ly$\alpha$                & 24.781 & 25$_{-8}^{+5}$    & 0$^{+600}_{-0}$        & -570$^{+240}_{-120}$ \\
23.51 $\pm$ 0.02        & \ion{O}{i}$_{Gal}$                     & 23.521 & 29$_{-8}^{+10}$   & 0$^{+940}_{-0}$        & -180$^{+240}_{-230}$ \\
22.06 $\pm$ 0.01        & \ion{O}{vii} f                         & 22.101 & -67 $\pm$ 15      & 980$^{+1070}_{-980}$   & -550 $\pm$ 160 \\
21.72 $\pm$ 0.03        & \ion{O}{vii} i+r$^{\mathrm{a}}$        &        & -68$_{-18}^{+16}$ & 2710$^{+1270}_{-790}$  & $-$ \\
18.88 $\pm$ 0.02        & \ion{O}{viii} Ly$\alpha$               & 18.969 & 24 $\pm$ 7        & 1240 $\pm$ 910         & -1370$^{+320}_{-360}$ \\
17.55 $\pm$ 0.03        & \ion{Fe}{xviii}                        & 17.591 & -9 $\pm$ 5        & 20$^{+1440}_{-20}$     & -770$^{+340}_{-460}$ \\
15.97 $\pm$ 0.01        & \ion{O}{viii} Ly$\beta$                & 16.006 & 30$_{-5}^{+6}$    & 960 $\pm$ 770          & -610 $\pm$ 230 \\
14.17 $\pm$ 0.01        & \ion{Fe}{xviii}                        & 14.210 & 23$_{-4}^{+8}$    & 0$^{+2400}_{-0}$       & -760$^{+170}_{-210}$ \\
13.431 $\pm$ 0.007      & \ion{Fe}{xix}/\ion{Ne}{ix} He$\alpha$  & 13.462$^{\mathrm{b}}$ & 61 $\pm$ 6 & 1710 $\pm$ 330 & -690 $\pm$ 160 \\
12.83 $\pm$ 0.01        & \ion{Fe}{xx}                           & 12.832 & 35 $\pm$ 8        & 1680$^{+930}_{-1240}$  & -160 $\pm$ 310 \\
\hline   
\end{tabular}
\begin{list}{}{}
\item[$^{\mathrm{a}}$] No laboratory wavelength or blueshift are listed for this feature, since it is a blend of lines whose apparent average wavelength is heavily dependent upon the physical conditions and geometry of the emitting region.
\item[$^{\mathrm{b}}$] Appropriately weighted by the relative abundances of iron and neon as listed by \citet{anders1989}.
\end{list}
\end{table*} 											

We identify the most prominent feature, significant at $\sim$ 16$\sigma$, with an absorption line blend of \ion{Fe}{xix} and \ion{Ne}{ix} He$\alpha$ at a blueshift of 690 $\pm$ 160 km~s$^{-1}$ (with the rest wavelength appropriately weighted for the relative abundances of iron and neon as listed by \citet{anders1989}). Several of the other lines correspond to transitions of H-like and He-like oxygen, nitrogen and carbon, and various states of sulphur, silicon and L-shell iron. The strongest lines (Table~\ref{eqws}), for which we can obtain the most reliable wavelength measurements, are mostly blueshifted by a few hundred km~s$^{-1}$. The blueshift of \ion{O}{viii} Ly$\alpha$ is much higher than this, probably as a result of the bias introduced by the presence of the emission part of the P-Cygni profile. There is also evidence for a higher velocity component, over 2000~km~s$^{-1}$, in the \ion{O}{vii} He$\alpha$ and \ion{C}{vi} Ly$\alpha$ lines (see Fig~\ref{vel_space}).

Four of the lines listed in Table~\ref{eqws} were also detected by \citet{scott2005}: \ion{O}{vii} f and \ion{O}{viii} Ly$\alpha$ emission and \ion{Ne}{ix} He$\alpha$ and \ion{Fe}{xx} $\lambda$12.832 absorption. The properties of these lines are broadly consistent between our analysis and theirs. The principal difference is that our FWHMs are less well-constrained due to the lower wavelength resolution of the RGS. The \ion{O}{viii} Ly$\alpha$ emission line was found to have a velocity shift consistent with zero by \citet{scott2005}, whereas it is blueshifted by 1370$^{+320}_{-360}$~km~s$^{-1}$ in our spectrum. The apparent velocity shift of this line is strongly influenced by the presence of \ion{O}{viii} Ly$\alpha$ absorption, so a change in the depth or outflow speed of the absorber could account for the discrepancy. The velocity shift of the \ion{O}{vii} f line, which is less affected by superimposed absorption, is more similar between the two analyses (-550 $\pm$ 160 in this analysis versus -200 $\pm$ 90~km~s$^{-1}$, which are probably consistent within the wavelength uncertainty of the RGS), so it seems most probable that the changing apparent velocity shift of \ion{O}{viii} Ly$\alpha$ is due to changing absorption properties. Our \ion{Fe}{xix}/\ion{Ne}{ix} He$\alpha$ absorption line has an equivalent width roughly equal to the combined equivalent widths of the two \ion{Ne}{ix} He$\alpha$ components observed by \citet{scott2005}. Of these two components, one is blueshifted by 670 $\pm$ 180~km~s$^{-1}$ and one is redshifted by 860 $\pm$ 100~km~s$^{-1}$; the velocity shift and FWHM of the blueshifted component are consistent with those for our \ion{Fe}{xix}/\ion{Ne}{ix} He$\alpha$ line.

\subsection{Ion$-$by$-$ion spectral modelling}

\begin{figure*}
   \centering
   \includegraphics[width=12cm]{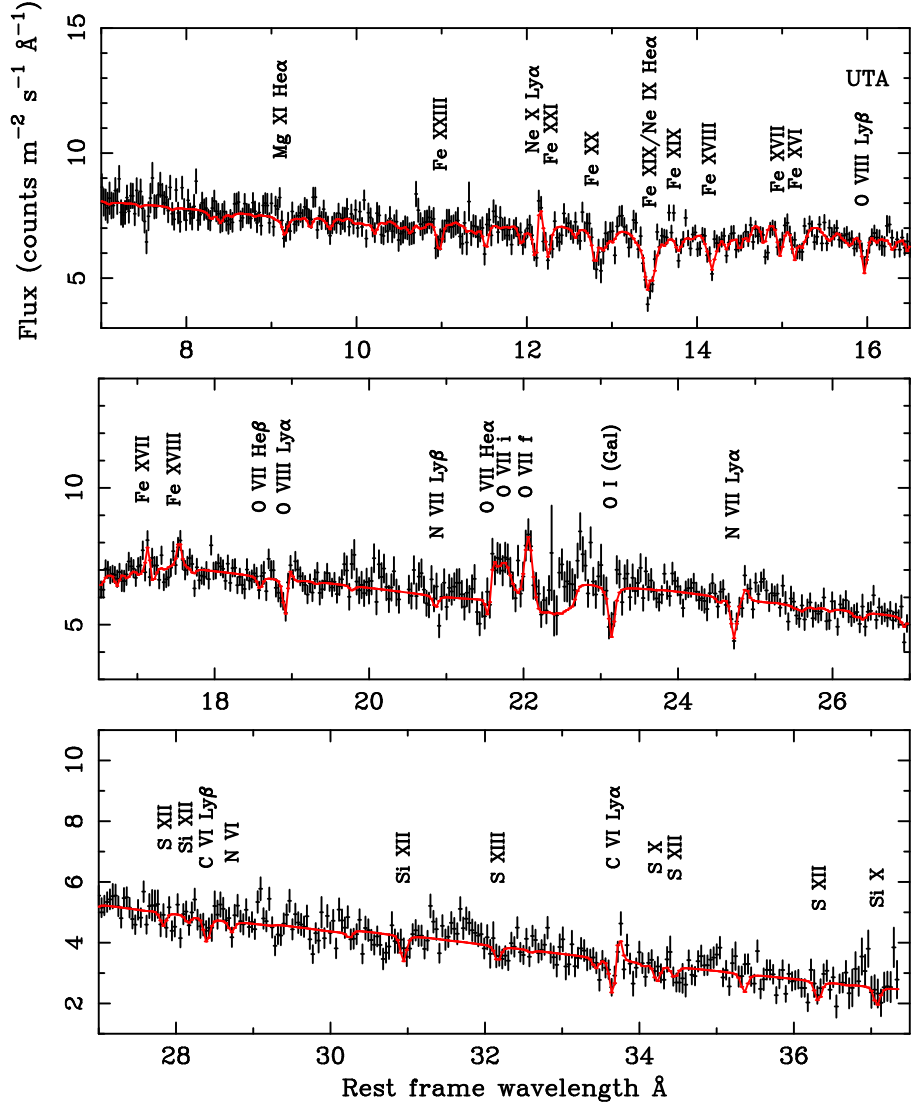}
      \caption{The best-fit \emph{slab} model overplotted on the RGS spectrum of NGC 7469, in the rest frame of the galaxy, with ionic species labelled.
              }
         \label{slab_model_plot}
   \end{figure*}

Taking the ions already identified as a starting point, we then used the \emph{slab} model in SPEX 2.00.11 \citep[][http://www.sron.nl/divisions/hea/spex/]{kaastra1996} to fit the absorbing columns of individual ions. We model the continuum underlying the RGS spectrum with a hard power-law and a soft black-body component, as seen in the pn spectrum; the parameters of this model are given in Table~\ref{slab_model}. We consider this to be a convenient parameterisation rather than a physical model; the nature of the soft excess in AGN is still under debate \citep[see~e.g.][]{brocksopp2006,crummy2006,gierlinski2006}. The velocity shift and turbulent velocity associated with the absorption lines were allowed to vary during the fit, and were assumed to be the same for all states. Since the deepest absorption lines for some of the most important species form part of P-Cygni profiles, it was also necessary to include the set of emission lines corresponding to these transitions; the parameters of these lines were fitted simultaneously with those of the corresponding absorption features. The parameters of the \emph{slab} best-fit model are listed in Table~\ref{slab_model}, and the fitted parameters of the emission lines are given in Table~\ref{slab_model_lines}. The model is shown overplotted on the RGS spectrum in Fig.~\ref{slab_model_plot}. The P-Cygni line profiles and very high fluxes of the narrow emission lines may already imply an extended flow and a significant mass outflow rate, as also suggested by \citet{behar2003} for \object{NGC~3783}.

\begin{table}
\caption{Parameters of the best-fit \emph{slab} model for the RGS spectrum of \object{NGC~7469}: the slope $\Gamma$ and normalisation $Norm_{PL}$ (in $10^{51}$~photons~s$^{-1}$~keV$^{-1}$) of the hard power-law; the temperature $t$ (keV) and normalisation $Norm_{BB}$ ($10^{16}$ m$^{2}$) of the soft blackbody; the average velocity shift $v_{shift}$ and RMS turbulent velocity $v_{turb}$ of the absorber (both in km~s$^{-1}$); the log of the column densities of ionic species in the spectrum in cm$^{-2}$.}
\label{slab_model}
\centering  
\begin{tabular}{c c}
\hline\hline 
$\Gamma$ & 1.81 $\pm$ 0.04  \\ 
$Norm_{PL}$ & 5.9 $\pm$ 0.1 \\ 
$t$  & 0.144$^{+0.007}_{-0.005}$ \\
$Norm_{BB}$  & 84$^{+30}_{-20}$ \\
\hline   
$v_{shift}$  & -690 $\pm$ 40 \\
$v_{turb}$  &  97 $\pm$ 9\\
\hline   
\ion{C}{vi}    &   16.7 $^{+0.2}_{-0.3}$     \\
\ion{N}{vi}    &   15.4 $^{+0.4}_{-1.4}$     \\
\ion{N}{vii}   &   16.5 $\pm$ 0.2	     \\
\ion{O}{vii}   &   16.3 $^{+0.2}_{-0.3}$     \\
\ion{O}{viii}  &  17.96 $^{+0.07}_{-0.08}$   \\
\ion{Ne}{ix}   &   17.2 $^{+0.2}_{-0.3}$     \\
\ion{Ne}{x}    &   17.3 $^{+0.3}_{-0.4}$     \\
\ion{Mg}{xi}   &   16.4 $^{+0.6}_{-0.7}$     \\
\ion{Mg}{xii}  &   16.5 $^{+0.5}_{-1.9}$     \\
\ion{Si}{x}    &   16.8 $^{+0.2}_{-0.4}$     \\
\ion{Si}{xi}   &   15.9 $^{+0.4}_{-1.1}$     \\
\ion{Si}{xii}  &   16.7 $\pm$ 0.2	     \\
\ion{S}{x}     &   16.2 $^{+0.3}_{-0.5}$     \\
\ion{S}{xii}   &   15.9 $\pm$ 0.3 	     \\
\ion{S}{xiii}  &   15.7 $^{+0.3}_{-0.6}$     \\
\ion{Fe}{ix}   &   15.7 $^{+0.2}_{-0.3}$     \\
\ion{Fe}{x}    &   15.8 $^{+0.1}_{-0.2}$     \\
\ion{Fe}{xi}   &   15.5 $^{+0.3}_{-0.7}$     \\
\ion{Fe}{xii}  &   15.0 $^{+0.5}_{-29}$      \\
\ion{Fe}{xiii} &   15.7 $\pm$ 0.2	     \\
\ion{Fe}{xiv}  &   15.4 $^{+0.3}_{-0.6}$     \\
\ion{Fe}{xv}   &   15.3 $^{+0.4}_{-1.2}$     \\
\ion{Fe}{xvi}  &   15.3 $^{+0.3}_{-1.0}$     \\
\ion{Fe}{xvii} &   16.0 $^{+0.2}_{-0.3}$     \\
\ion{Fe}{xviii}&   16.3 $\pm$ 0.1	     \\
\ion{Fe}{xix}  &   16.4 $\pm$ 0.1 	     \\
\ion{Fe}{xx}   &   16.3 $\pm$ 0.1	     \\
\ion{Fe}{xxi}  &   16.6 $\pm$ 0.2	     \\
\ion{Fe}{xxii} &   15.5 $^{+0.6}_{-30}$      \\
\ion{Fe}{xxiii}&   16.5 $^{+0.3}_{-0.4}$     \\
\hline   
\end{tabular}
\end{table} 					

\begin{table}
\caption{Parameters of the emission lines included with the \emph{slab} and \emph{xabs} model fits to the RGS spectrum of \object{NGC~7469}: line identification, $FWHM$ in km~s$^{-1}$, velocity shift $v_{shift}$ in km~s$^{-1}$ and line flux $F$ in $10^{-6}$ photons~cm$^{-2}$~s$^{-1}$.}
\label{slab_model_lines}
\centering  
\begin{tabular}{c c c c}
\hline\hline 
Line & $FWHM$ & $v_{shift}$  &  $F$ \\ 
\hline   
\ion{C}{vi} Ly$\alpha$   & 0$^{+2000}_{-0}$       & -70$^{+240}_{-470}$  &  30 $\pm$ 10  \\
\ion{N}{vii} Ly$\alpha$  & 0$^{+11000}_{-0}$      & 1100$^{+540}_{-270}$  & 9 $\pm$ 7   \\
\ion{O}{vii} f           & 1200$^{+1000}_{-700}$  & -520 $\pm$ 200  &  70 $\pm$ 10  \\
\ion{O}{vii} i           & 2500$^{+1500}_{-2500}$ & -760$^{+980}_{-720}$  &  60 $\pm$ 20  \\
\ion{O}{vii} r           & 970$^{+1800}_{-970}$   & -110$^{+1300}_{-670}$  &  40 $\pm$ 10 \\
\ion{O}{viii} Ly$\alpha$ & 640$^{+630}_{-640}$    & -520$^{+430}_{-230}$  & 80$^{+60}_{-30}$   \\
\ion{Ne}{x} Ly$\alpha$   & 0$^{+1000}_{-0}$       & 300$^{+440}_{-230}$  &  18 $\pm$ 6  \\
\ion{Fe}{xviii} $\lambda$17.56\AA\   & 0$^{+1900}_{-0}$ & -770 $\pm$ 310 &  15 $\pm$ 6  \\
\ion{Fe}{xvii} $\lambda$17.14\AA\    & 0$^{+5400}_{-0}$ & 1700$^{+740}_{-300}$  &  22 $\pm$ 5  \\
\hline   
\end{tabular}
\end{table} 											

%

\begin{figure}
   \centering
   \includegraphics[width=6cm,angle=-90]{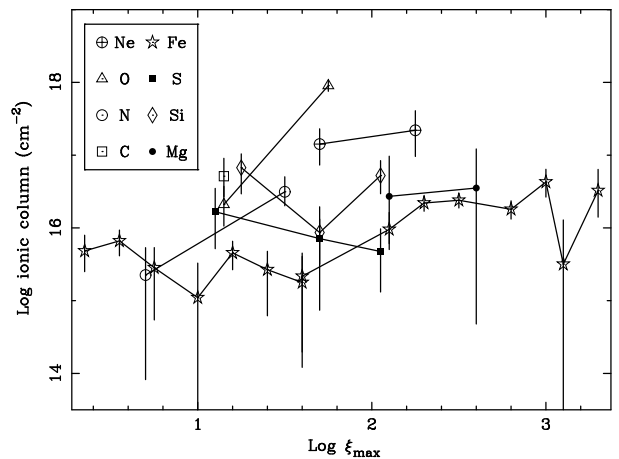}
      \caption{Plot of the absorbing columns of ions in the \emph{slab} model against the log ionisation parameter $\xi$ for which the ion is most abundant.
              }
         \label{slab_ions}
   \end{figure}

In Fig.~\ref{slab_ions}, the absorbing columns of the ions in this model are plotted against the ionisation parameter of greatest abundance of that ion. The plot displays the usual trend \citep[c.f.][]{blustin2005} of higher absorbing columns at higher levels of ionisation: this is due to an observational bias in that highly ionised gas needs to have a higher column in order to leave a visible trace in the spectrum. The lower-ionised states of iron fit the apparent Unresolved Transition Array (UTA) feature between 16$-$17~\AA\,. Unusually, there is no evidence of the lowly ionised oxygen states that generally accompany iron UTAs \citep[e.g.][]{blustin2002}. If these UTA ions are indeed present $-$ the \emph{slab} fit improves by a $\Delta\chi^{2}$ of 82 when the UTA ions are added to it $-$ then this may be an indication of a high relative abundance of iron in the lowest ionisation levels visible to us. The fitting of the UTA ions is also dependent upon the assumed outflow velocity, as the spectral resolution does not allow the identification of individual ions within it. The fitted RMS velocity of the gas, 97 $\pm$ 9 km~s$^{-1}$, corresponds to a Doppler width of $b$ = 137 $\pm$ 13 km~s$^{-1}$. This is close to the value inferred as most probable by \citet{scott2005}, who estimated $b$ $\sim$ 100 km~s$^{-1}$ for the soft X-ray absorption lines on the basis of the widths of nine-velocity-component UV absorption features measured in FUSE spectra.

%

\subsection{Absorption measure distribution}
\label{section_amd}

Fig.~\ref{slab_ions} shows that there is a broad range of ionisation present in the absorber, which spans about four orders of magnitude in $\xi$. Following the method of \citet{holczer2007}, we express the total hydrogen column density N$_H$ along the line of sight as an integral over its distribution in $\log \xi$. We term this distribution the Absorption Measure Distribution ($AMD = \partial N_H / \partial(\log\xi)$). The column density of each ion $N_{ion}$ can be expressed in terms of N$_{\rm H}$ as:

\begin{equation}
N_{ion} = A_z\int AMD f_{ion}(\log \xi)d(log\xi)
\end{equation}

\noindent where $A_z$ is the elemental abundance with respect to hydrogen, assumed to be constant throughout the absorber, and $f_{ion}$ is the fractional ion abundance with respect to the total abundance of its element, which we obtain from XSTAR \citep{kallman1982} using the SED described in Section~\ref{section_xabs}.

\begin{figure}
   \centering
   \includegraphics[width=7cm]{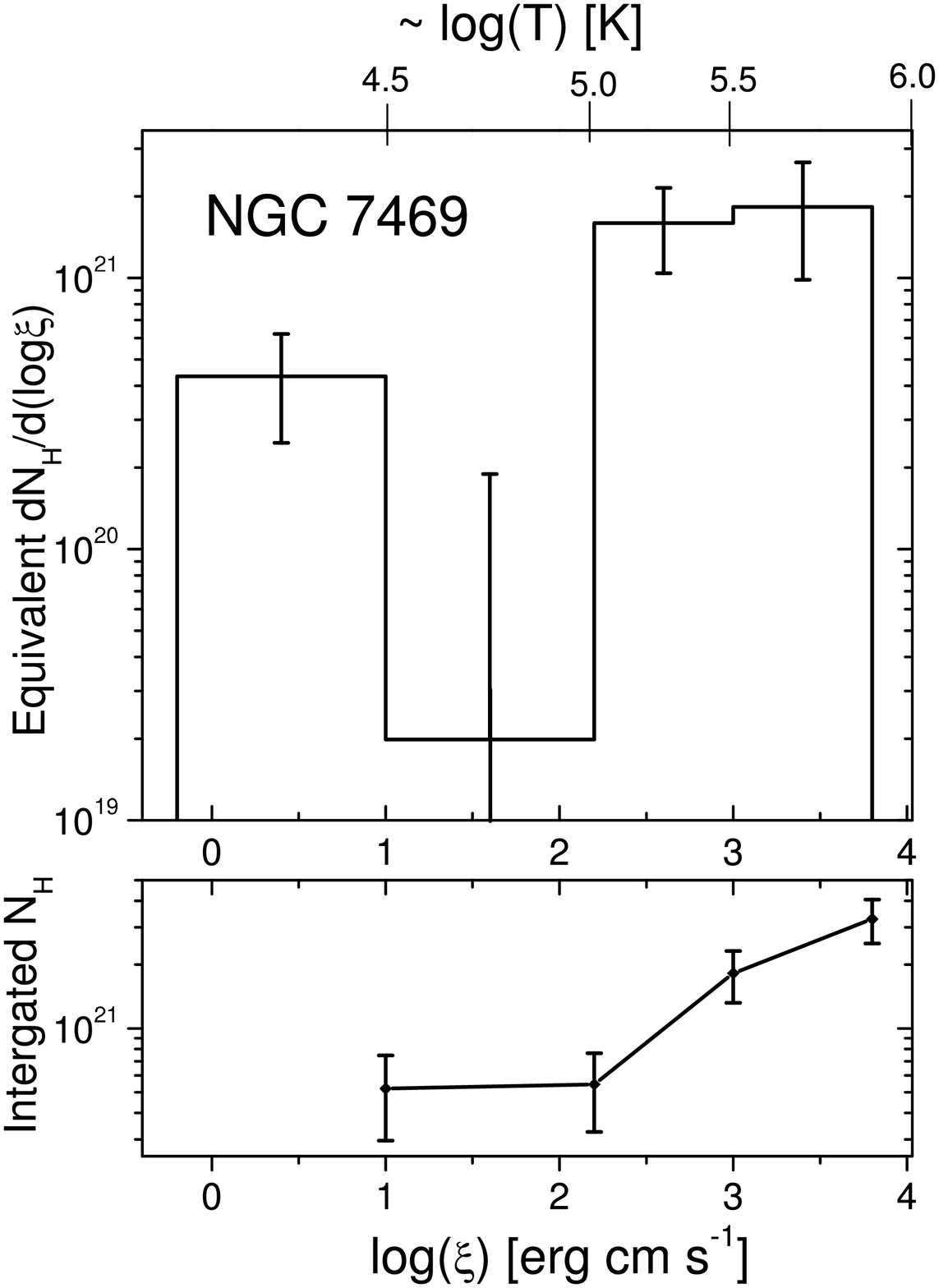}
      \caption{The Absorption Measure Distribution (AMD) of \object{NGC~7469} reconstructed from the ionic column densities of iron listed in Table~\ref{slab_model}. The AMD has a minimum at 4.5 K$< \log T < $ 5 K, demonstrating that the absorber is divided into two main phases. The integrated column density up to $\xi$ is plotted in the lower panel. The corresponding temperature scale is shown at the top of the figure. 
              }
         \label{amd}
   \end{figure}

We find the $AMD$ that best reproduces the whole set of Fe ionic column densities simultaneously. This distribution is presented in Fig.~\ref{amd}; we get a total equivalent hydrogen column density of $N_{\rm H}$ = (3.3$\pm$0.8)$\times10^{21}$~cm$^{-2}$. More details on the $AMD$ reconstruction method and the associated error computations can be found in \citet{holczer2007}. In Fig.~\ref{amd}, there is a minimum at $1.0 < \log\xi < 2$ corresponding to $4.5 < \log T <5$ (K), and a much higher column density in the high$-$ionisation component than in the low$-$ionisation component. The $AMD$ minimum at $4.5 < \log T <5$ (K) is seen in other AGN outflows as well \citep{holczer2007} and we speculate that it is due to a ubiquitous thermal instability in these winds. The abundances are obtained by rescaling the AMD (Fig.~\ref{amd}) to reproduce the measured non-Fe ionic column densities. The resulting values are listed in Table~\ref{amd_abund} and are broadly consistent with Solar.

%

\begin{table}
\caption{Abundances with respect to Solar \citep{anders1989} derived from the Absorption Measure Distribution (AMD.}
\label{amd_abund}
\centering  
\begin{tabular}{c c}
\hline
Element   & Abundance \\
\hline 
C  & 0.7$^{+0.9}_{-0.7}$  \\
N  & 0.6$^{+0.6}_{-0.4}$  \\
O  & 1.3 $\pm$ 0.5  \\
Ne & 3$^{+2}_{-1}$  \\
Mg & 1.0$^{+2}_{-0.7}$  \\
Si & 11$^{+9}_{-8}$  \\
S  & 6$^{+10}_{-5}$  \\
Fe & 1$^{\mathrm{a}}$     \\
\hline   
\end{tabular}
\begin{list}{}{}
\item[$^{\mathrm{a}}$] Fixed
\end{list}
\end{table}

\subsection{Multi$-$phase warm absorber model}
\label{section_xabs}

Taking the ionisation distribution revealed by the AMD as a starting point, we constructed a multi-phase spectral model of the warm absorber using the \emph{xabs} model in SPEX. \emph{Xabs} is based on a grid of XSTAR output, and generates absorption features due to photoionised gas at a given ionisation parameter and column density. The ionisation parameter $\xi$ is defined as 

   \begin{equation}
      \xi = \frac{L_{ion}}{n r^{2}} \,,
   \end{equation}
\label{xi}

in which $L_{ion}$ is the 1$-$1000~Rydberg ionising source luminosity (in erg s$^{\rm -1}$), $n$ the gas density (in cm$^{\rm -3}$) and $r$ the source distance in cm, so $\xi$ has the units erg cm s$^{\rm -1}$ \citep{tarter1969}.

\begin{figure}
   \centering
   \includegraphics[width=6cm,angle=-90]{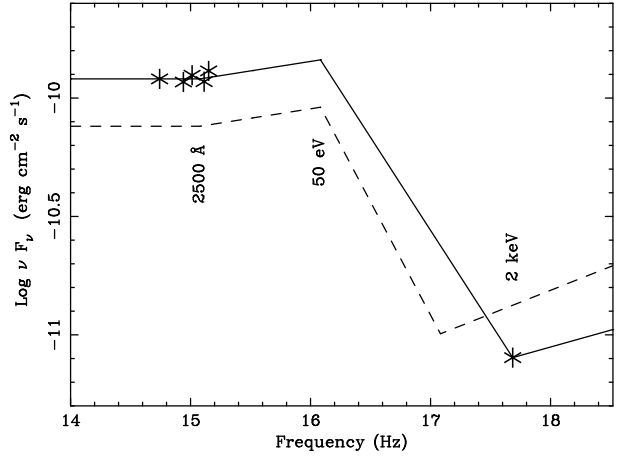}
      \caption{The \object{NGC~7469} spectral energy distribution at the time of the XMM-Newton observation in late 2004 (solid line), as inferred from the OM and pn monochromatic fluxes (stars). The SED used by \citet{scott2005} to represent data taken two years earlier is also plotted for comparison (dashed line).              }
         \label{sed}
   \end{figure}

The Spectral Energy Distribution (SED) that we used in calculating the \emph{xabs} models is based on that used by \citet{scott2005} to represent the ionising continuum of \object{NGC~7469} as it was in 2002, but adjusted for the broadband spectrum of the source as observed with XMM-Newton two years later. The hard power-law, above 2~keV, has $\Gamma=1.86$ to match the slope fitted to the EPIC-pn spectrum between 2$-$10~keV (with 6$-$7~keV ignored), and is normalised to the unabsorbed flux of the XMM-Newton spectrum at 2~keV. The soft excess begins to rise from $\sim$ 2~keV, so we take this energy to define the high-energy end of a power-law joining the UV and soft X-rays, and, following \citet{scott2005}, we set the low-energy end at 50~eV. Below this energy, we use the same spectral form as \citet{scott2005}, with $\Gamma=2$ below 2500~\AA\, and $\Gamma=1.92$ between 2500~\AA\, and 50~eV. The lowest-energy section was normalised to the V-band (5430~\AA) flux from the OM, and the resulting 50~eV$-$2~keV slope was calculated to be $\Gamma=2.78$. The spectrum cuts off below 0.1~eV and above 100~keV. Fig.~\ref{sed} is a plot of the resulting SED over the spectral range of interest, with monochromatic fluxes for the OM photometry points and the pn at 2~keV overplotted. The SED of \citet{scott2005} is also plotted for comparison.

%

\begin{figure*}
   \centering
   \includegraphics[width=12cm,angle=-90]{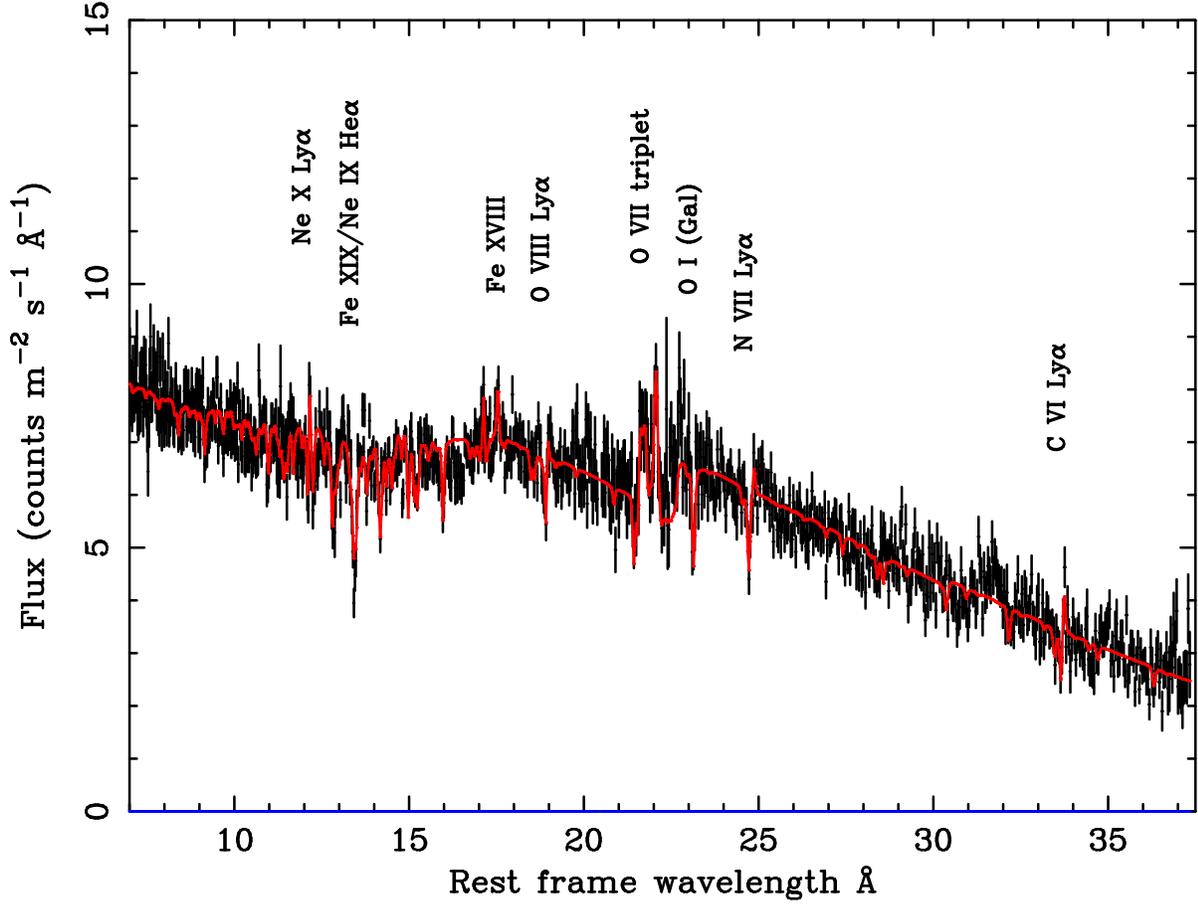}
      \caption{RGS spectrum composed of data from the whole observation with the three-phase spectral model superimposed: the model parameters are listed in Table~\ref{xabs_model}. 
              }
         \label{total_rgs_spec}
   \end{figure*}

The AMD shows (Fig.~\ref{amd}) that the absorber contains two principal ionisation ranges: a low ionisation phase with log $\xi$ $\sim$ 0.5 and a higher ionisation phase with an ionisation range of log $\xi$ $\sim$ 2.5$-$3.5. We constructed a spectral model with three \emph{xabs} components to reproduce this, with one component for the low ionisation gas and two components to represent the high ionisation range. The value for the turbulent velocity obtained in the \emph{slab} fit, 97~km~s$^{-1}$, was used for all phases, and the abundances were fixed at Solar \citep{anders1989}. The emission lines were included with parameters fixed at those previously fitted alongside the \emph{slab} model (see Table~\ref{slab_model_lines}). The absorbing columns, ionisation parameters and outflow speeds of the three phases obtained from the \emph{xabs} fits are listed in Table~\ref{xabs_model}. A plot of the spectrum with the model overlaid is given in Fig.~\ref{total_rgs_spec}. 

\begin{table}
\caption{Parameters of an \emph{xabs} model fit to the RGS spectrum of \object{NGC~7469}, using three phases of gas at different levels of ionisation. For each phase, the following parameters are listed: log ionisation parameter $\xi$ (erg cm s$^{-1}$), equivalent hydrogen column density $N_{H}$ (cm$^{-2}$), and average velocity shift $v_{shift}$ in km~s$^{-1}$.}
\label{xabs_model}
\centering  
\begin{tabular}{c c c}
\hline
Phase   & Parameter & Value \\
\hline 
Component 1 & Log $\xi$ & 0.8$^{+0.4}_{-0.3}$   \\ 
 & $N_{H}$ & 3$^{+2}_{-1}$ $\times$ 10$^{19}$    \\ 
 & $v_{shift}$  & -2300 $\pm$ 200    \\ 
\hline   
Component 2 & Log $\xi$ & 2.73 $\pm$ 0.03  \\ 
 & $N_{H}$ & 2.0 $\pm$ 0.2 $\times$ 10$^{21}$  \\ 
 & $v_{shift}$  & -720 $\pm$ 50  \\ 
\hline   
Component 3 & Log $\xi$ & 3.56$^{+0.08}_{-0.07}$ \\ 
 & $N_{H}$ & 2.9$^{+0.9}_{-0.7}$ $\times$ 10$^{21}$  \\ 
 & $v_{shift}$  & -580$^{+80}_{-50}$ \\ 
\hline   
\end{tabular}
\begin{list}{}{}
\item[$^{\mathrm{a}}$] Fixed
\end{list}
\end{table}

%

\begin{figure}
   \centering
   \includegraphics[width=7cm]{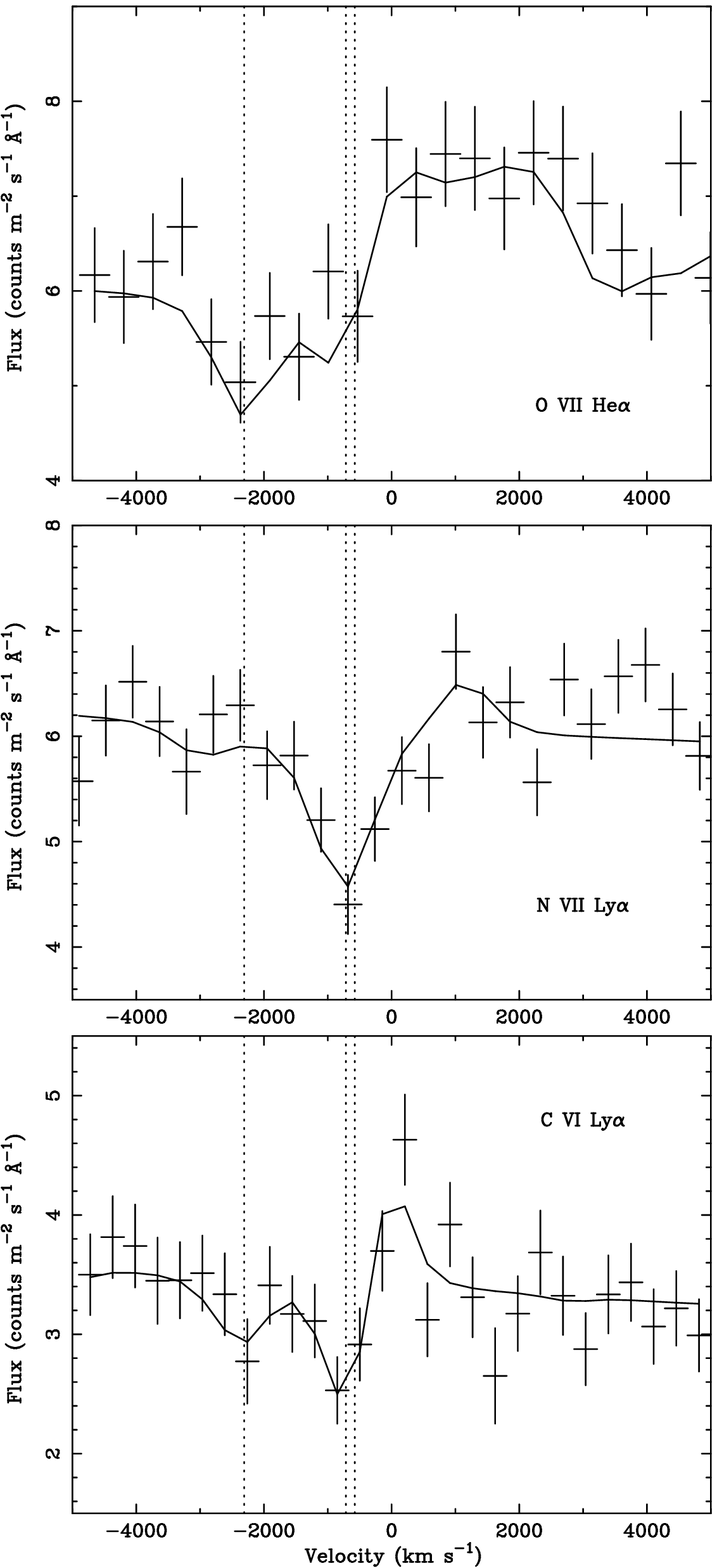}
      \caption{Velocity spectra of the \ion{O}{vii} He$\alpha$, \ion{N}{vii} and \ion{C}{vi} Ly$\alpha$ lines, showing the two principal velocity phases in the outflow. The dotted lines mark the fitted outflow speeds of the three \emph{xabs} components, at 2300, 720 and 580 km~s$^{-1}$ for components 1, 2 and 3 respectively (see Table~\ref{xabs_model}. 
              }
         \label{vel_space}
   \end{figure}

We found that whilst the higher ionisation phases have fitted outflow velocities of 720 and 580 km~s$^{-1}$, the lowest ionisation material was consistent with a higher velocity outflow phase at 2300 km~s$^{-1}$. Fig~\ref{vel_space} contains plots of the two velocity components associated with the \ion{O}{vii} He$\alpha$, \ion{N}{vii} and \ion{C}{vi} Ly$\alpha$ lines with the model overlaid. 

%

The low ionisation component 1 should give rise to absorption by states of M-shell iron forming an Unresolved Transition Array. With Solar elemental abundances, our model does not reproduce the full depth of this absorption; this is probably due to uncertainties in the ionisation balance of the UTA. A recent re-calculation of the dielectronic recombination rates for Fe$^{8+}-$Fe$^{12+}$ with a modern atomic physics code \citep{badnell2006} has indeed shown that the total recombination rates for these ions, radiative plus dielectronic, are an order of magnitude higher than those currently used in XSTAR and other photoionised plasma codes.

\section{Discussion}

\subsection{The mass-energy budget of the outflow}

When we want to quantify the mass transfer involved in AGN feedback in the context of galaxy evolution, how important is it to take the whole ionisation range of the outflow into account? To answer this, we need to know how much mass is coming out in all of the ionisation phases that we observe. We can estimate the mass outflow rates $\dot M$ of the UV and soft X-ray absorbing outflow phases in \object{NGC~7469} using Eqn.~18 in \citet{blustin2005}:

   \begin{equation}
      \dot M \sim \frac {1.23 m_p L_{ion} {C_v} v \Omega}{{\xi}}  \,,
   \end{equation}

where $m_p$ is the proton mass, $L_{ion}$ is the ionising (1$-$1000~Ryd; 13.6~eV$-$13.6~keV) luminosity, ${C_v}$ is the volume filling factor of a given ionisation phase, $v$ is the outflow speed, $\Omega$ is the solid angle subtended by the outflow and $\xi$ is the ionisation parameter (Eqn.~1). As in \citet{blustin2005}, we take $\Omega=1.6$, which is estimated from the observed fraction of type-1 AGN in the nearby universe, $\sim$ 25\% \citep{maiolino1995}, and the observation that the covering factor of warm absorber outflows in type-1 AGN is at least 50\% \citep{reynolds1997}. Note however that $\Omega$ could be a lot smaller than this in an outflow consisting of, for example, narrow streamers of gas \citep{steenbrugge2005}.

We can obtain a simple estimate of the volume filling factor ${C_v}$ if we assume that the outflow is radiatively accelerated. In such an outflow, the momentum of an outflowing phase (dependent upon ${C_v}$) must be of the order of the momentum of the radiation that it absorbs, $P_{abs}$, plus that of the radiation that it scatters, $P_{scatt}$. The resulting expression for $C_v$ is given by Eqn.~23 in \citet{blustin2005}:
   \begin{equation}
   {C_v} \sim \frac {{({\dot P_{abs}} + {\dot P_{scatt}})} {\xi}}{1.23 m_p L_{ion} v^2 \Omega} 
   \end{equation}
(note that the factor of $c$ in the denominator of this equation as originally printed was a typo). We can write
   \begin{equation}
      {\dot P_{abs}} = \frac {L_{abs}}{c}  \,,    
   \end{equation}
in which $c$ is the speed of light, and $L_{abs}$ is the luminosity absorbed by the outflow over the whole 1$-$1000~Ryd range; the absorbed luminosity is obtained using the \emph{xabs} warm absorber model in SPEX, taking the non-unity covering factors of the UV phases into account. Also
   \begin{equation}
      {\dot P_{scatt}} = \frac {L_{ion}}{c}(1 - e^{-\tau_{T}})  \,,   
   \end{equation}
where $\tau_{T}$ (the optical depth for Thomson scattering) is given by
   \begin{equation}
      \tau_{T} = \sigma_{T} {\rm N_{\rm H}} \,,
   \end{equation}
$\sigma_{T}$ is the Thomson cross-section and N$_{\rm H}$ is the absorbing column.

Having obtained the mass outflow rates, we can also calculate the kinetic luminosities of the outflow phases, using
   \begin{equation}
      L_{KE} = \frac{1}{2} {\dot M} v^2  \,.
   \end{equation}

To get an accurate figure for the total mass outflow rate, we also need to know whether any of the UV absorption is due to the same phases of gas that are absorbing in the X-rays. We can test whether this is the case by comparing low-ionisation ion columns predicted by our \emph{xabs} models with the columns actually measured using transitions in the UV spectra. In Table~\ref{uv_cols}, we list \ion{C}{iv}, \ion{N}{v} and \ion{H}{i} column densities predicted by our models alongside values obtained from the June 2004 STIS spectra \citep{scott2005}, which are the measurements closest in time to our December 2004 X-ray spectra. We also list the ionisation parameters and velocities of the two UV and three X-ray phases; the UV phase ionisation parameters were derived using spectra taken in 2002, so we have rescaled them to take account of the greater ionising luminosity at the later epoch ($2.19\times10^{44}$ in 2004 versus $1.45\times10^{44}$~erg~s$^{-1}$ in 2002). The velocities of the UV phases did not change significantly between the 2002 and 2004 observations; using the components of the Ly$\alpha$ absorption line as an example, the column-weighted average high-velocity component was -1896 $\pm$ 3~km~s$^{-1}$ in 2002 and -1893 $\pm$ 4~km~s$^{-1}$ in 2004, whilst the low-velocity component was -560 $\pm$ 10~km~s$^{-1}$ and -580 $\pm$ 10~km~s$^{-1}$ in 2002 and 2004 respectively.

   \begin{table*}
    
      \caption[]{A comparison of some properties of the X-ray and UV absorbing outflow phases in \object{NGC~7469}: absorber phase, log ionisation parameter (log $\xi$; erg cm s$^{\rm -1}$), outflow speed ($v$; km~s$^{\rm -1}$), \ion{C}{iv} column density ($10^{14}$ cm$^{\rm -2}$), \ion{N}{v} column density ($10^{14}$ cm$^{\rm -2}$), \ion{H}{i} column density ($10^{14}$ cm$^{\rm -2}$).}
         \label{uv_cols}  
   \centering
         \begin{tabular}{lllllr}
            \hline
            \noalign{\smallskip}
Phase & Log $\xi$ & $v$ & N$_{\ion{C}{iv}}$ & N$_{\ion{N}{v}}$ & N$_{\ion{H}{i}}$  \\
            \noalign{\smallskip}
            \hline
            \noalign{\smallskip}
X-ray component 1 & 0.8$^{+0.4}_{-0.3}$    & -2300 $\pm$ 200    & 1.6             & 3.4           & 6.2           \\
X-ray component 2 & 2.73 $\pm$ 0.03        & -720 $\pm$ 50      & Negligible      & 0.00091       & Negligible    \\
X-ray component 3 & 3.56$^{+0.08}_{-0.07}$ & -580$^{+80}_{-50}$ & Negligible      & Negligible    & Negligible    \\
UV component 1 & 1.61$^{\mathrm{a}}$       & -562$^{\mathrm{b}}$ $\pm$ 6       & 0.98 $\pm$ 0.09 & 2.9 $\pm$ 0.8 & 7 $\pm$ 2     \\
UV component 2 & 0.51$^{\mathrm{a}}$       & -1901$^{\mathrm{b}}$ $\pm$ 6      & 2.0 $\pm$ 0.1   & 2.5 $\pm$ 0.2 & 2.4 $\pm$ 0.5 \\
            \noalign{\smallskip}
            \hline
         \end{tabular}        
\begin{list}{}{}
\item[$^{\mathrm{a}}$] Converted from U to $\xi$ using the \citet{scott2005} SED, and rescaled to take into account the change in ionising luminosity between the 2002 and 2004 observations.
\item[$^{\mathrm{b}}$] As measured from 2002 FUSE and STIS data \citep{scott2005}.
\end{list}
  \end{table*}

The high-speed, low ionisation X-ray component 1 has an ionisation parameter consistent with that of UV component 2. Its outflow velocity is similar, especially taking the systematic wavelength uncertainty of the RGS (at 100$-$200~km~s$^{-1}$) into account. The columns of \ion{C}{iv}, \ion{N}{v} and \ion{H}{i} predicted to be present in this phase are very close to the actual values measured in UV spectra. We therefore identify UV component 2 with X-ray component 1. Of the other ionisation phases, the outflow speeds of X-ray components 2 and 3 may be consistent with that of UV component 1, but the ionisation levels of the phases, and predicted columns, are not consistent. This is the opposite conclusion to that reached by \citet{blustin2003} and \citet{kriss2003}, namely that the high-ionisation phase was identical with UV component 1, but UV component 2 lacked an X-ray analogue. We note, however, that our new X-ray spectra are vastly superior to the original observations of \citet{blustin2003}.

   \begin{table*}
    
      \caption[]{Physical properties for each warm absorber phase in \object{NGC~7469}: absorber phase, log ionisation parameter (log $\xi$; erg cm s$^{\rm -1}$), log column density (log $N_{\rm H}$; cm$^{\rm -2}$), outflow speed ($v$; km~s$^{\rm -1}$), covering factor ($f$), percentage volume filling factor (\% $C_{\rm v}$), mass outflow rate per phase (${\dot M}_{\rm out}$; M$_\odot$ yr$^{\rm -1}$), log kinetic luminosity of outflow (log $L_{KE}$; erg s$^{\rm -1}$).}
         \label{mdotcomp}  
   \centering
         \begin{tabular}{llllllllllr}
            \hline
            \noalign{\smallskip}
Phase & Log $\xi$ & Log N$_{\rm H}$ & $v$ & $f$ & \% $C_{\rm v}$ & ${\dot M}_{\rm out}$ & Log $L_{KE}$ \\
            \noalign{\smallskip}
            \hline
            \noalign{\smallskip}
X-ray component 1 & 0.8  & 19.5 & 2300 & 1 & 0.0005 & 0.002 & 39.6  \\
X-ray component 2 & 2.73 & 21.3 & 720  & 1 & 2      & 0.03  & 39.7  \\
X-ray component 3 & 3.56 & 21.5 & 580  & 1 & 12     & 0.02  & 39.4  \\
UV component 1    & 1.61$^{\mathrm{a}}$ & 20.0 & 562  & 0.53 & 0.04    & 0.006  & 38.7  \\
UV component 2    & 0.51$^{\mathrm{a}}$ & 18.6 & 1901 & 0.93 & 0.00007 & 0.0004 & 38.7  \\
            \noalign{\smallskip}
            \hline
         \end{tabular}        
\begin{list}{}{}
\item[$^{\mathrm{a}}$] Converted from U to $\xi$ using the \citet{scott2005} SED, and rescaled to take into account the change in ionising luminosity between the 2002 and 2004 observations.
\end{list}
  \end{table*}

Table~\ref{mdotcomp} gives the observed parameters and calculated percentage volume filling factors, mass outflow rates, and kinetic luminosities for the X-ray and UV absorbing phases. Taking all of the X-ray and UV phases into account, but assuming that UV component 2 is identified with X-ray component 1 (and using the mass outflow rate of the X-ray phase, which is higher), we obtain a total mass outflow rate of 0.06 M$_\odot$ yr$^{\rm -1}$. Calculated in the same way, the total kinetic luminosity of the outflow is 1.2 $\times$ 10$^{40}$ erg~s$^{-1}$. The X-ray absorbing gas is responsible for $\sim$ 90\% of the mass outflow rate and 95\% of the kinetic luminosity of the outflow. 

In the case of UV component 2/X-ray component 1, the mass outflow rate and kinetic luminosity calculated from the UV properties are respectively 21\% and 14\% those derived from the X-ray properties. This is partly due to the slightly higher outflow velocity measured from the X-ray spectrum (resulting from the larger wavelength uncertainty of the X-ray measurements), but also to the order of magnitude lower column density obtained from the UV. What we have not yet tested is how much mass and energy is transported by matter at much higher ionisation levels, containing ionisation states that primarily absorb at energies above 6~keV. This will be easiest to do for high-column warm absorbers, like for instance \object{NGC~3783}, where there will be more absorption features visible at the highest energies \citep[see~e.g.][]{reeves2004}.

We note here that although our estimates are based on the assumption of a radiatively-accelerated outflow, the nature of the acceleration mechanism in AGN winds remains controversial. Models of radiatively-driven winds in AGN have the difficulty that the gas can become so highly ionised that the optical depth is no longer great enough for radiative acceleration to take place. This problem can be overcome where parts of the outflow are shielded from the soft X-rays by either a highly ionised `blanket' layer, or so-called hitch-hiker gas which is caught up in the accelerating outflow \citep[see~e.g.][]{murray1995}, a rather more complex scenario than the simplified analysis we present here. Other models involve the gas being thermally driven \citep[e.g.][]{krolik2001} or flowing out from an accretion disc along (locally) open magnetic field lines \citep[e.g.][]{bottorff2000}. In some models this magnetically launched gas is then radiatively accelerated \citep[e.g.][]{konigl1994,everett2005}. \citet{everett2006} and \citet{das2006} attempt to model the observed outflowing Narrow Line Region (NLR) dynamics in nearby Seyfert 2s using, respectively, thermally and radiatively driven winds. Neither model could reproduce the acceleration profile of the outflowing clouds, and it seems likely that interaction with the surrounding medium has a significant effect on the dynamics of AGN winds.

\subsection{The ionised outflow in \object{NGC~7469}: a torus wind being launched?}

We can use the measured properties of the warm absorber in \object{NGC~7469} to estimate the distances of the different absorbing phases from the central engine. Upper and lower limits for these distances can be obtained following Eqns.~27 and 32 of \citet{blustin2005}. The minimum distances are obtained on the assumption that the outflowing gas must have reached the escape velocity, and the maximum distance from the criterion that the gas must be close enough for it to be ionised. The minimum distance is likely to be the weaker constraint. We may be observing the gas before it has reached its final velocity, so it could be closer to the central engine than we predict. Observations of outflowing gas in the NLRs of local Seyfert 2s shows that the wind accelerates up to a distance of order 100~pc \citep{ruiz2005,das2005,das2006}, so we can certainly say that it has effectively escaped from the gravitational potential well of the black hole. It does not escape from the \emph{galaxy's} potential well, however, as these same observations show that the gas decelerates again by the time it reaches a distance of 300$-$400~pc. This deceleration is very likely, however, to be at least partly the result of interactions with the surrounding medium.

\begin{figure}
   \centering
   \includegraphics[width=6cm,angle=-90]{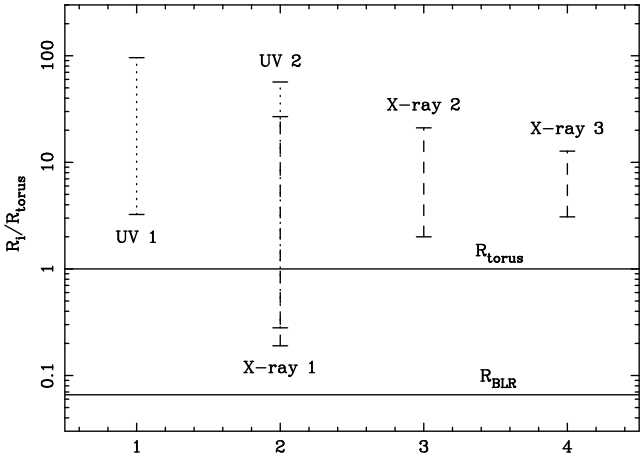}
      \caption{Locations of the X-ray (dashed lines) and UV (dotted lines) absorbing phases with respect to the BLR and torus. Y-axis: upper and lower bounds to the distances of the absorbing phases from the central engine, expressed as ratios to the distance of the torus. The ratio of the BLR to torus distances is also plotted (X-axis is arbitrary units). 
              }
         \label{phase_dists}
   \end{figure}

Table~\ref{wa_dists} lists the calculated distance ranges, and Fig.~\ref{phase_dists} shows these distances plotted as ratios to the distance of the torus. Using dust reverberation, \citet{suganuma2006} show that the inner radius of the torus in \object{NGC~7469} is in the range of 65$-$87 light-days from the nucleus. For our purposes, we take the torus distance to be the average of these values, which corresponds to 0.064~pc. The location of the BLR is also plotted in Fig.~\ref{phase_dists}, where we used the reverberation mapping distance of 5 light-days measured by \citet{wandel1999}.

   \begin{table}
    
      \caption[]{Distance ranges of the ionised outflow phases from the central engine (R, pc).}
         \label{wa_dists}  
   \centering
         \begin{tabular}{llllllllllr}
            \hline
            \noalign{\smallskip}
Phase &  R \\
            \noalign{\smallskip}
            \hline
            \noalign{\smallskip}
X-ray 1          & 0.012$-$1.7  \\
X-ray 2          & 0.13$-$1.3  \\
X-ray 3          & 0.20$-$0.81  \\
UV component 1 (2004 epoch) & 0.21$-$6.1  \\
UV component 2 (2004 epoch) & 0.018$-$3.6 \\
            \noalign{\smallskip}
            \hline
         \end{tabular}        
  \end{table}

%

With the estimates we obtain here, we find that all of the phases are further out than the BLR, and most are further out than the torus. This contradicts the previous finding that UV component 1, which was identified with the high ionisation X-ray absorber, is probably at or within the BLR \citep{kriss2003,blustin2003}. This inconsistency was first noticed by \citet{scott2005}, who also applied the distance estimates presented by \citet{blustin2005}, and discussed a number of reasons for the disagreement. The main evidence for the X-ray/UV component 1 absorber phases being located at or within the BLR was that UV component 1 has a very low covering factor of $\sim$ 50\%, consistent with a partial covering of both the broad lines and continuum, or covering only the UV continuum and not the broad lines. Its relatively high ionisation parameter and column, which associated it with the main body of the X-ray absorber, supported this conclusion. Component 2 has a much higher covering factor of $\sim$ 90\%, which implies that it is further away from the central engine. The faster outflow speed of this more distant component indicates that the wind is accelerating.

The new information we can now add to this, with much better quality X-ray data, is that the low ionisation X-ray absorption is likely to be identified with the high velocity UV component 2, due to the similar outflow velocities and predicted column densities of low-ionisation ionic states observable in the UV. The high ionisation phases of the X-ray absorber are unlikely to be identified with UV component 1, principally because the ionic columns measured earlier in 2004 using STIS are not predicted by our \emph{xabs} models for these phases, even though the velocities of the phases are probably consistent (considering the 7~m\AA\, absolute wavelength uncertainty of the RGS, which is equivalent to $\sim$100~km~s$^{-1}$ at 20\AA). 

The low covering factor of UV component 1 could be consistent with our distance estimates (Table~\ref{wa_dists}, Fig.~\ref{phase_dists}) if it comprises a filamentary structure, perhaps near the base of a wind where there is acceleration of clumpy, inhomogeneous matter from the dusty torus. The very low volume filling factor we derive for this phase, at 0.04\%, supports this interpretation. The X-ray absorber, at a similar speed to component 1, would also be located close to the base of the outflow, where one would expect gas at a range of ionisation levels to exist as matter is ionised and driven outwards. UV component 2 probably represents gas which has been accelerated away from the base of the outflow, as it is moving faster, and its higher covering factor indicates greater distance. The X-ray and UV properties of the \object{NGC~7469} warm absorber may, therefore, be the signature of a torus wind in the process of being launched.

\section{Conclusions}

We use our $\sim$ 160~ks soft X-ray spectrum of \object{NGC~7469}, the highest quality soft X-ray data yet obtained for this source, to obtain accurate measurements of the ionisation state, absorbing column and outflow speed of its warm absorber. We show that there is gas over a wide range of ionisation (log $\xi$ $\sim$ 0.5$-$3.5), and we find evidence for two separate velocity regimes at 580$-$720 and 2300~km~s$^{-1}$. Contrary to expectations from previous observations, the properties of the high ionisation X-ray absorber are not consistent with those of either of the components of the UV absorber, but the lowest ionisation X-ray absorber is probably identical with the highest velocity UV phase. We estimate the distances of the X-ray and UV absorbing phases from the central engine, finding that they are most consistent with a location close to the torus or between the torus and BLR, and discuss a scenario where the covering factors and velocities of these absorber phases signify the launch and acceleration of a wind from the torus.

Using the observed properties of the X-ray absorbing phases, along with the luminosity-scaled properties of the UV absorber obtained from observations two years earlier \citep{scott2005}, we calculate the mass outflow rates and kinetic luminosities of all of the soft X-ray and UV warm absorber phases, taking into account the identification of an X-ray phase with a UV phase. We estimate that $\sim$ 90\% of the mass outflow rate and $\sim$ 95\% of the kinetic luminosity are associated with the X-ray absorbing phases. A complete picture of the mass/energy budget of the multi-ionisation-phase warm absorber also needs to include the highest ionisation gas, which primarily absorbs above 6~keV. The properties of this high-ionisation gas are most easily measured in outflows with higher absorbing columns than \object{NGC~7469}.

For this nearby AGN, we have demonstrated that the mass and energy outputs are probably dominated by the X-ray warm absorber, although an appreciable fraction is only detectable in the UV band. It will be interesting to see how far this conclusion extends to AGN with cosmologically-interesting BAL-type outflows, but it is clear that any observational study of AGN feedback via ionised winds needs to take the whole ionisation range of the outflow into account.

\begin{acknowledgements}

Based on observations obtained with XMM-Newton, an ESA science mission with instruments and contributions directly funded by ESA Member States and NASA. The UCL-MSSL authors acknowledge the support of PPARC. The research at the Technion was supported by ISF grant 28/03 and by a grant from the Asher Space Research Institute. The Netherlands Institute for Space Research (SRON) is supported financially by NWO, the Netherlands Organisation for Scientific Research. SK acknowledges financial support by the Zeff fellowship at the Technion. We thank the referee, E. Mediavilla, for useful comments.

\end{acknowledgements}

\end{document}